\documentclass[letterpaper, 10 pt, conference]{ieeeconf}  

\IEEEoverridecommandlockouts                              

\usepackage{graphics} 
\usepackage{epsfig} 
\usepackage{amsmath} 
\usepackage{amssymb}  
\usepackage{bm}
\usepackage{subcaption}
\usepackage{algorithm}
\usepackage{algorithmicx}
\usepackage{algpascal}
\newtheorem{remark}{Remark}
\title{\LARGE \bf Kernel-Based Identification of Local Limit Cycle Dynamics with \\ Linear Periodically Parameter-Varying Models}

\author{Defne E. Ozan$^{1}$, Mingzhou Yin$^{2}$, Andrea Iannelli$^{2}$ and Roy S. Smith$^{2}$
\thanks{This work was partially supported by the Swiss National Science Foundation under Grant 200021\_178890.}
\thanks{$^{1}$The author is with the Department of Aeronautics, Imperial College London, SW7 2AZ, United Kingdom,
{\tt\small d.ozan@imperial.ac.uk}.
}
\thanks{$^{2}$The authors are with the Department of Electrical Engineering, Automatic Control Laboratory, ETH Z\"{u}rich, 8092 Z\"{u}rich, Switzerland,
{\tt\small myin/iannelli/rsmith@control.ee.ethz.ch}.
}
}

\begin{document}

\maketitle
\thispagestyle{empty}
\pagestyle{empty}

\noindent
\fbox{\begin{minipage}{\linewidth}

D. E. Ozan, M. Yin, A. Iannelli and R. S. Smith, ``Kernel-Based Identification of Local Limit Cycle Dynamics with Linear Periodically Parameter-Varying Models," arXiv:2203.16306.

\vspace{0.5em}

\textcopyright\ 2022 IEEE. Personal use of this material is permitted. Permission from IEEE must be obtained for all other uses, in any current or future media, including reprinting/republishing this material for advertising or promotional purposes, creating new collective works, for resale or redistribution to servers or lists, or reuse of any copyrighted component of this work in other works.
\end{minipage}}
\vspace{1em}

\begin{abstract}
Limit cycle oscillations are phenomena arising in nonlinear dynamical systems and characterized by periodic, locally-stable, and self-sustained state trajectories. Systems controlled in a closed loop along a periodic trajectory can also be modelled as systems experiencing limit cycle behavior. The goal of this work is to identify from data, the local dynamics around the limit cycle using linear periodically parameter-varying models. Using a coordinate transformation onto transversal surfaces, the dynamics are decomposed into two parts: one along the limit cycle, and one on the transversal surfaces. Then, the model is identified from trajectory data using kernel-based methods with a periodic kernel design. The kernel-based model is extended to also account for variations in system parameters associated with different operating conditions. The performance of the proposed identification method is demonstrated on a benchmark nonlinear system and on a simplified airborne wind energy model. The method provides accurate model parameter estimation, compared to the analytical linearization, and good prediction capability.\\
\end{abstract}

\section{Introduction}
Nonlinear dynamical systems of two or higher dimensions can exhibit periodic solutions known as \textit{limit cycle oscillations} \cite{Strogatz1994}. Limit cycles are isolated closed orbits that if locally stable, are local attractors, and thus lead to self-sustained periodic oscillations. When a system is controlled along a periodic reference, the closed-loop dynamics can also be considered a limit cycle.
In this regard, it is of interest to identify a model that describes the dynamics of limit cycles, which can then be used for simulation, analysis, and control design. Nevertheless, identification of nonlinear systems purely from data poses a difficult problem, which requires prior knowledge of the model structure, and/or complex nonlinear optimization schemes with tractability issues \cite{Schoukens2019}. Instead, local linear dynamics are often identified for different operating points to construct a linear parameter-varying (LPV) model and apply gain scheduling in control design \cite{toth2010modeling}. For limit cycles, oftentimes the local dynamics are of main concern. However, conventional LPV methods do not take into account that the underlying model converges to a limit cycle. In comparison, linearization of the system directly around the limit cycle as in \cite{Allen2009} results in a linear time-varying (LTV) model. This model fails to capture the dynamics along the limit cycle, i.e. the velocity at which the perturbed trajectories traverse the points on the limit cycle while converging to it. In this study, an alternative approach that identifies the nonlinear dynamics around the limit cycle as a linear periodically parameter-varying (LPPV) model is investigated. The first step is to decompose the dynamics into two parts: one moving along the limit cycle, and one lying on the transversal hyperplanes of the limit cycle, which are known as Poincar\'e sections. This decomposition implies a transformation onto the so-called \textit{transverse coordinates} \cite{Manchester2011}. Next, the dynamics around the limit cycle are modelled as a periodic system parametrized with the location on the limit cycle. In the vicinity of the limit cycle, the system can be approximated with a locally linearized model. LPPV modelling bridges the gap between existing LPV and LTV approaches for the periodic case.

The linearized transverse dynamics reduce the identification problem to learning the system matrices as functions of the location on the limit cycle, which are periodic in nature. In this work, this function learning problem is tackled by using kernel methods in an LPV system identification framework \cite{Bachnas2014}. Conventional parametric approaches require a priori selection of an appropriate set of basis functions that map the given inputs to a higher dimensional nonlinear feature space. Kernel methods allow this mapping to be done implicitly onto an infinite-dimensional function space and the function can be estimated in this function space with Tikhonov regularization. Such methods have been previously used for the nonparametric identification of LPV systems in \cite{Laurain2012} in an input-output setting, and in \cite{Rizvi2018} with state-space models. This work extends the method proposed in \cite{Rizvi2018} with a separate kernel design for each element of the system matrices, and the periodicity in the learned system matrices is enforced via periodic kernel design. In addition, the flexibility of kernel design makes it possible to include additional system parameters in the model, by augmenting the periodic kernel with standard non-periodic kernels. 

The algorithm is first tested on the Van der Pol oscillator. The identified model is demonstrated to be close to analytical linearization when training data are close to the limit cycle, and outperform analytical linearization in terms of prediction accuracy when the training data are close to the prediction task. Then, the algorithm is applied to a simplified kinematic model of a tethered kite controlled to fly along a periodic figure-of-eight trajectory for airborne wind energy generation \cite{Diehl2013}. Accurate prediction can be obtained with an additional system parameter. The proposed method performs significantly better than global nonlinear identification without knowledge of the limit cycle.

\section{Transverse Dynamics of Limit Cycles}\label{sec:transverse}
In this section, the background of transverse dynamics of limit cycles is summarized. For detailed definitions and derivations, see \cite{Hale1980,Manchester2011}. 

Let us consider a nonlinear system described by a set of ordinary differential equations (ODEs):
\begin{equation}\label{eq:nonlinear}
	\dot{x} = f(x,d),
\end{equation}
where $x \in \mathbb{R}^{n}$ is the state vector and $d \in \mathbb{R}^{n_d}$ is an exogenous input. The autonomous solution of this system, i.e., $\dot{x} = f(x,0)$, starting from an initial condition $x(0) = x_0$ is denoted by $x(t) = \Phi(x_0,t)$. The system exhibits limit cycle behaviour if it has a $T^\star$-periodic solution $x^\star(t) = \Phi(x^\star_0,t)$, i.e., $T^\star > 0 $ is the minimum period such that the relationship $x^\star(t) = x^\star(t+T^\star)$ holds for all $t$. Then, the limit cycle is defined as $\Gamma^\star = \{x\in \mathbb{R}^{n}: x = x^\star(\tau)|\tau \in [0,T^\star)\}$, where it is parametrized with a new time scale $\tau \in [0,T^\star)$. In this study, we consider asymptotically stable periodic orbits. The periodic orbit $\Gamma^\star$ is said to be asymptotically stable if it fulfills Lyapunov stability, i.e., $\forall\epsilon > 0$, $\exists\delta > 0$ such that $\forall x_0 \in \mathbb{R}^n$ with $\text{dist}(x_0, \Gamma^\star) < \delta$, $\text{dist}(\Phi(x_0,t), \Gamma^\star) < \epsilon$, $\forall t > 0$, and if it is an attractor, $\exists\delta > 0$ such that $\forall x_0 \in \mathbb{R}^n$ with $\text{dist}(x_0, \Gamma^\star) < \delta$, $\lim_{t\rightarrow\infty}\text{dist}(\Phi(x_0,t), \Gamma^\star)=0$, where $\text{dist}(x,\Gamma^\star) = \text{inf}_{y\in\Gamma^\star}||y-x||_2$. The disturbance $d$ is assumed to be such that the perturbed trajectories remain close to the nominal limit cycle. At each $\tau$, one can construct an $(n-1)$-dimensional hyperplane $S(\tau)$ that is transversal to $\Gamma^\star$, i.e., $\dot{x}^\star(\tau) \notin S(\tau)$. The transversal hyperplanes are uniquely defined by normal vectors denoted by $z(\tau)$. On this hyperplane, a new coordinate system is defined such that the origin is $x^\star(\tau)$ and the coordinate axes can be chosen as any orthonormal basis that spans the surface $S(\tau)$. The coordinates of a given state $x \in S(\tau)$ in this new coordinate frame are denoted by $x_\perp \in \mathbb{R}^{n_{\perp}}$, where $n_{\perp} = n-1$. Thus, a mapping of the state to its \textit{transverse coordinates} is created for a given family of transversal surfaces moving along the periodic orbit: $x\rightarrow(x_\perp,\tau)$. The collection of the basis vectors of $S(\tau)$ defines a projection operator $\Pi(\tau) = [\xi_1\ \dots\ \xi_{n-1}]^\top$ that characterizes the transformation to the transverse coordinates:
\begin{equation}\label{eq:map_x_perp_x}
x = x^\star(\tau) + \Pi(\tau)^\top x_\perp,
\end{equation}
and the inverse relationship is
\begin{equation}\label{eq:map_x_x_perp}
x_\perp = \Pi(\tau)(x-x^\star(\tau)).
\end{equation}

The transversality condition can be rewritten in terms of the normal vector as requiring the existence of some $\delta > 0$ such that $z(\tau)^\top \dot{x}^\star(\tau) > \delta$, $\forall\tau \in [0,T^\star)$. The most straightforward choice of surfaces is then those that are orthogonal to the orbit, i.e., the normal vectors are set to be tangential to the flow as
\begin{equation}
    z^\text{orth}(\tau) = \frac{\dot{x}^\star(\tau)}{||\dot{x}^\star(\tau)||_2}.
\end{equation}
However, this choice leads to singularities that occur especially around $\tau$ sections where the curvature of the orbit is large \cite{Manchester2011}. These singularities are due to the violation of the so-called well-posedness condition that arises from the nonlinear $\tau$ dynamics. This condition restricts the region where the transformation to transverse coordinates is well-defined.
An alternative set of surfaces is considered, originally proposed in \cite{Ahbe2022}. These surfaces, referred to as center surfaces, connect $x^\star(\tau)$ with a fixed center (e.g., the geometric center of the limit cycle). The first basis vector $\xi_1$ is
\begin{equation}\label{eq:center}
\xi_1^\text{center}(\tau) = \frac{x^\star (\tau)-x_c}{||x^\star (\tau)-x_c||_2}, \\
\end{equation}
where $x_c$ represents the designated center point. The remaining basis vectors can be selected such that the angle between the center surface and the orthogonal surface is the smallest. The normal vector $z^\text{center}(\tau)$ can be consequently determined as the unit vector perpendicular to all the basis vectors.

Subsequent to the relationships established in \eqref{eq:map_x_perp_x} and \eqref{eq:map_x_x_perp}, the dynamics of the transverse states can be analytically obtained. We are interested in the local linearized model of the transverse system of the form
\begin{subequations}
\begin{align}
	\dot{x}_\perp &= A(\tau)x_\perp + B(\tau)d
	\label{eq:linear_x_perp},\\
	\dot{\tau} &= 1 + g(\tau)x_\perp + h(\tau)d
	\label{eq:linear_tau},
\end{align}
\label{eq:linear_model}%
\end{subequations}
where $A(\tau): [0,T^\star) \rightarrow \mathbb{R}^{n_\perp \times n_\perp}$, $B(\tau): [0,T^\star) \rightarrow \mathbb{R}^{n_\perp \times n_d}$, $g(\tau): [0,T^\star) \rightarrow \mathbb{R}^{1 \times n_\perp}$, and $h(\tau): [0,T^\star) \rightarrow \mathbb{R}^{1 \times n_d}$ are periodically-varying matrix functions of $\tau$. When the nonlinear model is known, the system matrices can be obtained by analytical linearization of the transverse dynamics \cite{Manchester2011}. Note that if $x$ is on the limit cycle, i.e., $x_\perp = 0$, $\tau$ would be equal to $t$ when no exogenous input is applied. Otherwise, the $\tau$ dynamics would differ from $t$ and the transverse model encapsulates this behavior. In contrast, the LTV approach in \cite{Allen2009} results in the following model which ignores the $\tau$ dynamics \eqref{eq:linear_tau}:
$
    \dot{\tilde{x}} = \tilde{A}(t)\tilde{x} + \tilde{B}(t)d, 
$
where $\tilde{x}(t) = x(t)-x^\star (t)$.

\section{Identification of Linear Periodically Parameter-Varying Models}\label{sec:identification}
To simplify the notation, define $\theta =  \left[x_\perp^\top\ \  d^\top\right]^\top\in\mathbb{R}^{n_\theta}$, where $n_\theta = n_\perp+n_d$, and $\zeta =  \left[\dot{x}_{\perp}^\top\ \ \dot{\tau}-1\right]^\top \in \mathbb{R}^n$. The dynamics \eqref{eq:linear_model} can then be compactly rewritten as:
\begin{equation}\label{eq:linear_compact}
    \zeta = \Omega(\tau)\theta,
\end{equation}
where
\begin{equation}
    \Omega(\tau)=
    \begin{bmatrix}
        A(\tau)&B(\tau)\\
        g(\tau)&h(\tau)
    \end{bmatrix}:[0,T^\star) \rightarrow \mathbb{R}^{n \times n_\theta}.
\end{equation}

Assume that measurements of the original state trajectories, their time derivatives, and the exogenous inputs are given as $\{x(t_k),\dot{x}(t_k), d(t_k)\}_{k = 1}^{N}$, and the periodic orbit $\Gamma^\star$ is known. To convert a state $x$ to its transverse counterpart $(x_\perp,\tau)$, the corresponding hyperplane must first be determined. The problem can be reformulated as finding the $\tau$ that satisfies the hyperplane equation and minimizes the distance between $x$ and the corresponding point on the limit cycle:
\begin{equation}\label{eq:optimization}
	\begin{aligned}
		\underset{\tau} \min \quad &||x-x^\star(\tau)||_2, \\
		\text{s.t.}\quad &z(\tau)^\top (x-x^\star(\tau)) = 0.
	\end{aligned}
\end{equation}
This optimization problem is solved for each $\tau(t_k)$ by a nonlinear solver initialized from $\tau(t_{k-1})$.
The transverse coordinates $x_\perp$ are then computed using the projection in \eqref{eq:map_x_x_perp}. Finally, the time derivatives of the transverse states $(\dot{x}_\perp(t_k),\dot{\tau}(t_k))$ can be calculated from $\dot{x}$ using the nonlinear analytical expressions from Theorem~1 in \cite{Manchester2011}. Thus, the dataset $\{\theta(t_k),\zeta(t_k),\tau(t_k)\}_{k = 1}^{N}$ is obtained. 
\subsection{Kernel-Based Identification}
A natural approach to function learning problems is to assume that the underlying function can be decomposed into a set of continuous basis functions:
\begin{equation}\label{eq:weighted_bf}
    \Omega_i(\tau) = \sum_{m = 1}^{n_\psi}w_m^i \psi_m^i(\tau) = W_i \Psi_i(\tau)
\end{equation}
where $\Omega_i(\tau)$ denotes the $i$-th row of $\Omega(\tau)$, $\psi_m^i(\tau): [0,T^\star) \rightarrow \mathbb{R}^{1\times n_\theta}$ represent the preselected vector-valued basis functions, $w_m^i\in \mathbb{R}$ are the associated weights, and
\begin{equation}
     \Psi_i(\tau)=\left[\psi_1^i(\tau)^\top\ \dots\ \psi_{n_\psi}^i(\tau)^\top\right]^\top\!\!\!,\ W_i = \left[w_1^i\ \dots\ w_{n_\psi}^i\right]
\end{equation}
collects the basis functions and the weights respectively. Here, each row $\Omega_i(\tau)$ of the system matrix is considered separately and solved independently. In machine learning practices, such transformations are referred to as \textit{feature maps}.

The learning problem is then posed as a regularized least-squares problem:
\begin{equation} \label{eq:id_kernel_opt}
	\min_{W_i} \quad \sum_{k = 1}^{N}\left( \zeta_i(t_k)-W_i\Psi_i(\tau(t_k))\theta(t_k)\right)^2+\lambda_i ||W_i||_2^2,
\end{equation}
where a Tikhonov regularization with the weighting factor $\lambda_i \in \mathbb{R}$ is applied. The predictions of state derivatives $\zeta_i$ is denoted as
\begin{equation}\label{eq:id_kernel_compact}
    \hat{\zeta}_i = W_i\Psi_i(\tau)\theta.
\end{equation}
Problem \eqref{eq:id_kernel_opt} can be solved directly. However, the process of selecting the basis functions is not trivial and the dimension $n_\psi$ is typically very large. Instead, the kernel method is used to reformulate the problem. In detail, by formulating the dual problem of \eqref{eq:id_kernel_opt}, it is shown that the optimal solution of the weights $W_i$ lies in the span of the training data \cite{Rizvi2018}, \cite{Toth2011}:
\begin{equation}
	 W_i = \sum_{k = 1}^{N}\alpha_{i,k} \theta(t_k)^\top \Psi_i(\tau(t_k))^\top,
	 \label{eq:w_to_alpha}
\end{equation}
where $\alpha_{i,k} \in \mathbb{R}$ are the coefficients associated with each training point. The predicted $\zeta_i$ can thus be expressed as
\begin{equation}\label{eq:id_kernel_expanded}
	\hat{\zeta}_i(t_{k'}) = \sum_{k = 1}^{N}\alpha_{i,k}\theta(t_k)^\top \Psi_i(\tau(t_k))^\top \Psi_i(\tau(t_{k'}))\theta(t_{k'}).
\end{equation}
Then, problem \eqref{eq:id_kernel_opt} can be reformulated in terms of $\alpha_i = [\alpha_{i,1}\ \alpha_{i,2}\ \dots\ \alpha_{i,N}]^\top$, which only depends on the inner product of the feature map $K_i(\tau,\tau'):=\Psi_i(\tau)^\top\Psi_i(\tau')\in[0,T^*)\times[0,T^*)\rightarrow \mathbb{R}^{n_\theta \times n_\theta}$ instead of $\Psi_i(\tau)$. This inner product function $K_i(\cdot,\cdot)$ is known as the kernel, which can be conceptually thought of as a similarity measure between two data points. Since $n_\psi$ is usually much larger than $n_\theta$, one can directly design $K_i$ instead of $\Psi$ to avoid explicitly choosing maps and implicitly work with features of higher or infinite dimensions. The idea of replacing inner products of feature maps with kernels is known as the \textit{kernel trick} \cite{SchoelkopfBook2002}. Substituting the kernel into \eqref{eq:id_kernel_expanded}, we obtain
\begin{equation}\label{eq:id_kernel_final}
	\hat{\zeta}_i(t_{k'}) = \sum_{k = 1}^{N}\alpha_{i,k} \theta(t_k)^\top K_i(\tau(t_k),\tau(t_{k'}))\theta(t_{k'}).
\end{equation}
Assuming that the elements of the system matrices can be modelled independently from each other, the kernel functions $K_i$ are designed as diagonal matrices, i.e., $K_i = \text{diag}\left(k_{i,1},k_{i,2},\dots,k_{i,n_\theta}\right)$, where scalar kernels $k_{i,j}: [0,T^\star) \times [0,T^\star) \rightarrow \mathbb{R}$ are designed for each system matrix element $\Omega_{i,j}$. This kernel design generalizes \cite{Rizvi2018} where the same kernel is used for each element, i.e., $K_i = k_i \mathbb{I}_{n_\theta}$.
\begin{remark}
The matrix-valued kernel function $K_i(\cdot,\cdot)$ can also be directly designed as a full matrix to model correlations between the elements in $\Omega_i$ \cite{Alvarez2012}. However, this is beyond the scope of this paper.
\end{remark}

Then, the predictions on all training points can be expressed as
$
	\hat{Z}_i = [\hat{\zeta}_i(t_1)\ \hat{\zeta}_i(t_2)\ \dots\ \hat{\zeta}_i(t_N)]^\top = \Upsilon_i\alpha_i ,
$
where $\Upsilon_i\in\mathbb{R}^{N \times N}$ is a positive semi-definite matrix, whose $(k,k')$-th element is constructed as
\begin{equation}
	\left(\Upsilon_{i}\right)_{k,k'} = \theta(t_k)^\top K_i(\tau(t_k),\tau(t_{k'})) \theta(t_{k'}).
\end{equation}
Define the collection of state derivative measurements as $Z_i = \left[\zeta_i(t_1)\ \zeta_i(t_2)\ \dots\ \zeta_i(t_N)\right]^\top$. The solution to problem \eqref{eq:id_kernel_opt} can then be indirectly given by the closed-form solution of $\alpha_i$:
\begin{equation}
	\alpha_i = (\Upsilon_i+\lambda_i \mathbb{I}_N)^{-1}Z_i,
	\label{eq:alpha}
\end{equation}
through the transformation \eqref{eq:w_to_alpha}.
Finally, the system matrices are retrieved as
\begin{equation}\label{eq:id_kernel_sol}
	\Omega_i(\tau) = \sum_{k = 1}^{N} \alpha_{i,k} \theta(t_k)^\top K_i(\tau(t_k),\tau).
\end{equation}
\begin{remark}
The learned system matrix function \eqref{eq:id_kernel_sol} can also be interpreted as the solution to the regularized function learning problem within the reproducing kernel Hilbert space associated with the kernel $K_i(\cdot,\cdot)$, denoted by $\mathcal{H}_{K_i}$ \cite{SchoelkopfBook2002}:
\begin{equation}
	\min_{\Omega_i\in\mathcal{H}_{K_i}} \ \sum_{k = 1}^{N}\left( \zeta_i(t_k)-\Omega_i(\tau(t_k))\theta(t_k)\right)^2+\lambda_i ||\Omega_i||_{\mathcal{H}_{K_i}}^2.
\end{equation}
\end{remark}
\subsection{Periodic Kernel Design}
Since the system matrices are periodic, the periodic kernel design first proposed in \cite{Mackay1998} will be used for $k_{i,j}$. Periodic kernels of period $T^\star$ are constructed by applying the warping $\chi(\tau)=\left[\sin(\frac{2\pi}{T^\star}\tau)\ \cos(\frac{2\pi}{T^\star}\tau)\right]^\top$ to any standard kernel. We consider the squared exponential (SE) kernel described by
\begin{equation}
	k_{i,j}^{\text{SE}}(\tau,\tau') = \exp\left(-\frac{||\tau-\tau'||_2^2}{2l_{i,j}^2}\right),
\end{equation}
where $l_{i,j}$ are the hyperparameters, known as the length scale, which control the smoothness of the functions to be learned.
The corresponding periodic kernel is then obtained by substituting $\tau$ with $\chi(\tau)$ and rearranging using trigonometric identities:
\begin{equation}\label{eq:per_se_kernel}
	k_{i,j}^{\text{PSE}}(\tau,\tau') =\exp\left(-\frac{2\sin^2(\frac{\pi}{T^\star}(\tau-\tau'))}{l_{i,j}^2}\right).
\end{equation}
Note that for any $\tau-\tau'=mT^*$, $m\in\mathbb{Z}$, $k_{i,j}^{PSE}(\tau,\tau')=1$. This means that the function values at $\tau$ and $\tau'$ are perfectly correlated, so the functions learned with such kernels are periodic with period $T^\star$. 
\subsection{Extension to the Multivariate Case}\label{sec:multivar}
The above identification method can be extended to the case where the system is operated around different operating points, such that the dynamics are also parameter varying with a parameter $p$:
\begin{equation}\label{eq:nonlinear_p}
    \dot{x} = f(x,d;p).
\end{equation}
In terms of the transverse dynamics, \eqref{eq:nonlinear_p} implies an additional dependence on $p$ for the limit cycle $x^\star(\tau,p)$ and the linearized model $\zeta = \Omega(\tau,p)\theta$.
The kernel method provides a straightforward way to incorporate such dependence in identification. Multivariate functions can be learned by multiplying kernels \cite{Rasmussen2006}. In our case, to model the dependence on $p$, the periodic kernel can be multiplied with an SE kernel:
\begin{equation}\label{eq:multi_kernel}
	k^{\text{Multi}}\left(
	\begin{bmatrix}
	    \tau\\p
	\end{bmatrix},
	\begin{bmatrix}
	    \tau'\\p'
	\end{bmatrix}
	\right) = k^{\text{PSE}}(\tau,\tau')k^{\text{SE}}(p,p').
\end{equation}
\subsection{Hyperparameter Selection}
The empirical Bayes, or the maximum marginal likelihood approach, is used to identify the hyperparameters in the kernel method \cite{Rasmussen2006}, which are the length scales $l_i=\left[l_{i,1}\ \dots\ l_{i,n_\theta}\right]^\top\in\mathbb{R}^{n_\theta}$, associated with each kernel and the regularization parameters $\lambda_i$:
\begin{equation}
    \underset{l_i,\lambda_i}{\text{max}}\ \log p(Z_i|\{\theta(t_k),\tau(t_k)\}_{k=1}^N,l_i,\lambda_i),
    \label{eq:hyper_est}
\end{equation}
where the log marginal likelihood function is given by
\begin{equation}
\begin{split}
	\log p(Z_i|&\{\theta(t_k),\tau(t_k)\}_{k=1}^N,l_i,\lambda_i) =\\ &-\frac{1}{2}(Z_i^\top\bar{\Upsilon}_i^{-1} Z_i -\text{log det}\bar{\Upsilon}_i)-\frac{N}{2}\log(2\pi),
\end{split}
\end{equation}
where $\bar{\Upsilon}_i = \Upsilon_i+\lambda_i \mathbb{I}_N$.

\vspace{0.5em}

The proposed identification algorithm is summarized in Algorithm~\ref{al:1}.
\begin{algorithm}[tb]
	\caption{Kernel-based identification of local limit cycle dynamics with LPPV models}
	\begin{algorithmic}[1]
	\State \textbf{Input:} training data $\{x(t_k),\dot{x}(t_k), d(t_k)\}_{k = 1}^{N}$, limit cycle $\Gamma^\star$.
	\State Select transversal surfaces $S(\tau)$ and corresponding projection operators $\Pi(\tau)$.
	\State Find $\left\{x_\perp(t_k),\tau(t_k)\right\}_{k = 1}^{N}$ by \eqref{eq:optimization} and \eqref{eq:map_x_x_perp}.
	\State Find $\left\{\dot{x}_\perp(t_k),\dot{\tau}(t_k)\right\}_{k = 1}^{N}$ by Theorem~1 in \cite{Manchester2011}.
	\For{i:=1}{n}
	\Begin
	\State Find $l_i,\lambda_i$ by solving \eqref{eq:hyper_est} with kernel design \eqref{eq:per_se_kernel}.
	\State Find $\Omega_i(\tau)$ by \eqref{eq:alpha} and  \eqref{eq:id_kernel_sol}.
	\End
	\State \textbf{Output:} transverse system matrix $\Omega$.
	\end{algorithmic}
	\label{al:1}
\end{algorithm}

\section{Numerical Examples}\label{sec:numerical}
\subsection{Van der Pol System}\label{sec:vanderpol}
The nonlinear benchmark system known as the Van der Pol oscillator is described by: 
\begin{subequations}
\begin{align}
	\dot{x}_1 &= x_2, \\
	\dot{x}_2 &= \mu(1-x_1^2)x_2-x_1+D\sin(\omega t),
\end{align}
\label{eq:vdp_ss}%
\end{subequations}
where a sinusoidal forcing term corresponds to the external input $d$ in \eqref{eq:linear_model}. The damping coefficient $\mu$ is set to 1, which results in a limit cycle with period $T^\star = 6.663$.

\begin{figure}[tb]
	\centering
		\includegraphics[width=\columnwidth]{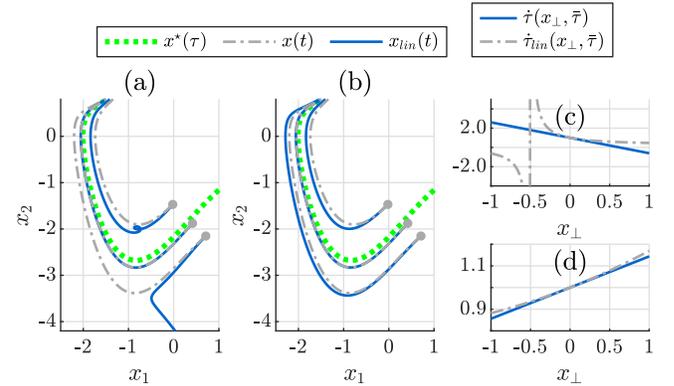}
	\caption{Effects of surface selection. Trajectory simulations in (a)-(b) and $\tau$ dynamics around a sharp turn in (c)-(d) using orthogonal (a),(c) and center (b),(d) surfaces.}
	\label{fig:vdp_surfaces}
\end{figure}
In Figure \ref{fig:vdp_surfaces}, nonlinear trajectories generated from \eqref{eq:vdp_ss} with $D = 0$ , denoted by $x(t)$, are compared to those obtained from the analytical transverse linear approximation $x_{lin}(t)$ using (a) orthogonal, and (b) center surfaces (the center point is chosen as the origin). For orthogonal surfaces, the well-posedness condition is violated around the sharp turns where the surfaces clash into each other, which causes a discontinuity in the nonlinear $\tau$ dynamics (Figure \ref{fig:vdp_surfaces}(c)). Around these regions, the transverse linear dynamics become unstable for large $x_\perp$ values (Figure \ref{fig:vdp_surfaces}(a)). This behavior is prevented by center surfaces, in which the linear dynamics $\dot{\tau}_{lin}$  can effectively approximate $\dot{\tau}$ (Figure \ref{fig:vdp_surfaces}(d)). These conclusions prompt the use of center surfaces for identification purposes.

Two sets of data, $\mathcal{D}_1$ and $\mathcal{D}_2$, are generated for identification, which contain trajectories starting from $x_\perp(t_0) = 0.1$ and $x_\perp(t_0) = -0.5$, respectively. For both sets, the forcing term is set as $D = 1$ and $\omega = 10\omega^\star$, and zero-mean Gaussian noise with a signal-to-noise ratio (SNR) of 40 dB is injected to state and state time-derivative measurements. The computation time is around $4\,$s in this example (on an Intel Core i7-9750H processor at 2.60GHz), which is dominated by the hyperparameter search step. 

\begin{figure}[tb]
\centering
    \includegraphics[width=\columnwidth]{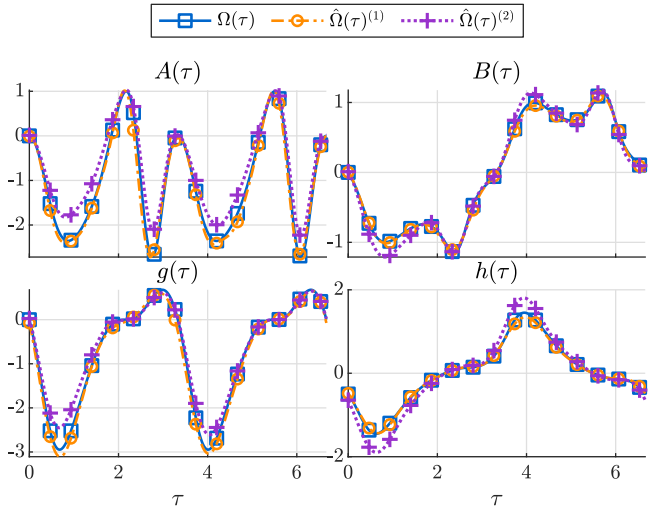}
    \caption{Comparison of the identified LPPV models for the Van der Pol system using different training datasets. $\Omega(\tau)$: analytical model, $\hat{\Omega}(\tau)^{(1)}$, $\hat{\Omega}(\tau)^{(2)}$: identified models using $\mathcal{D}_1$ and $\mathcal{D}_2$ respectively.}
    \label{fig:vdp_id_A_g}
\end{figure}

\begin{figure}[tb]
    \includegraphics[width=\columnwidth]{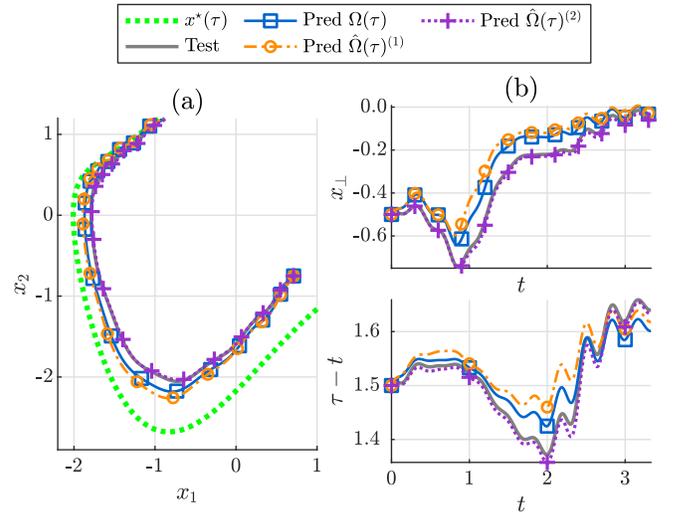}
    \caption{Prediction of a test trajectory from the Van der Pol system, shown in (a) the phase space and in (b) the time series of $x_\perp$ and $(\tau-t)$.}
    \label{fig:vdp_id_traj_pred}
\end{figure}

Figure \ref{fig:vdp_id_A_g} displays the identified system functions from $\mathcal{D}_1$ and $\mathcal{D}_2$, denoted by $\hat{\Omega}(\tau)^{(1)}$ and $\hat {\Omega}(\tau)^{(2)}$ respectively, alongside the analytical transverse linear system functions derived from the nonlinear system ODE, $\Omega(\tau)$. For $\mathcal{D}_1$, the identified model matches the analytical one linearized around $x_\perp = 0$. Predictions on a test trajectory with $x_\perp(t_0) = -0.5$, $\tau(t_0) = 1.5$, $D = 0.5$, $\omega = 20\omega^\star$ are shown in Figure \ref{fig:vdp_id_traj_pred} in (a) the phase space, and (b) time series plots of $x_\perp$ and $(\tau-t)$. By observing that $\hat{\Omega}(\tau)^{(2)}$ outperforms the other models in terms of prediction error, it can be concluded that the performance of the identification improves when the training data is chosen based on the regions in which the predictions are to be made, and can even be superior to an analytical linearization with a known nonlinear model. 
\subsection{Airborne Wind Energy System}\label{sec:awe}
A tethered kite system with ground-based power generation during the traction phase is investigated as a physical system model. The position of the kite is expressed by the elevation angle $\theta$, the azimuth angle $\phi$, and the line length $r$. The unicycle kinematic model from \cite{Wood2015} is considered:
\begin{subequations}
\begin{align}\label{eq:kite_ode}
	\dot{\theta} &= \frac{v}{r}\cos(\gamma),\\
	\dot{\phi} &= \frac{v}{r\cos(\theta)}\sin(\gamma), \\
	\dot{\gamma} &= u.
\end{align}
\end{subequations}
where $x = [\theta\ \phi\ \gamma]^\top$ is the state variable and $u$ is the steering input channel. 
The parameters $v$ and $r$ are assumed to be constant over one cycle.
The kite is controlled on an efficient figure-of-eight path by setting
	$\gamma^\star (\tau) = a\cos(\omega^\star \tau+b)$,
where the frequency $\omega^\star $, the amplitude $a$, and the phase $b$ are determined from the desired midpoint angles and system dynamics \cite{Wood2015}. The control law is designed as transverse state-feedback following \cite{Manchester2011,Ahbe2018}:
\begin{equation}\label{eq:transverse_feedback_input}
	u(\tau) = u^\star (\tau)+u_\perp(\tau) = u^\star (\tau)-K^\star (\tau)x_\perp(\tau).
\end{equation}
The nominal control input $u^\star(\tau)$ 
and the controller gains $K^\star (\tau)$ can be computed off-line and a periodically time-varying LQR controller is designed using the linearized periodic system matrix $A(\tau)$. The associated periodic differential Riccati equation \cite{Bittanti1991} is solved with the one-shot algorithm \cite{Varga2007}. 
The center surfaces have been defined starting from the first basis vector $\xi_1$ \eqref{eq:center}. 
The second basis vector is chosen as the vector perpendicular to both the first basis vector and the flow direction at that point given by $\dot{x}^\star(\tau)$.  

\begin{figure}[tb]
\centering
\includegraphics[width=\columnwidth]{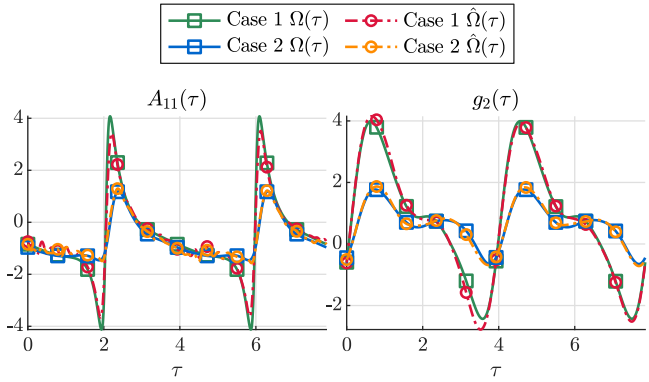}
\caption{Identified LPPV models for the tethered kite system with $\frac{v}{r}$ parametrization. Case 1: $\frac{v}{r} = 0.11$, Case 2: $\frac{v}{r} = 0.27$.}
\label{fig:kite_var}
\end{figure}
\begin{figure}[tb]
\centering
\includegraphics[width=0.95\columnwidth]{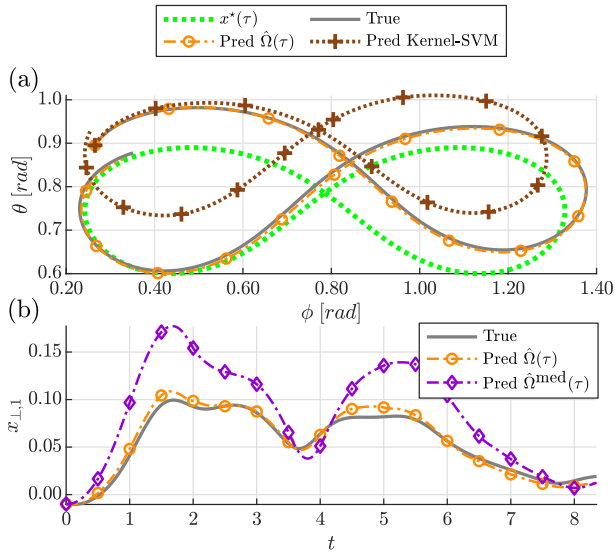}
\caption{Prediction of a test trajectory from the tethered kite system for $\frac{v}{r} = 0.27$, shown in (a) the phase space of $\theta$ and $\phi$, and in (b) the time series of $x_{\perp,1}$. Pred $\hat{\Omega}(\tau)$: identified multivariate model, Pred $\hat{\Omega}^{\text{med}}(\tau)$: identified model without $\frac{v}{r}$ parametrization.}
\label{fig:kite_var_traj_pred}
\end{figure}
The kite system is simulated with $\omega^\star = 0.8, \: \theta^\star(0) = \frac{\pi}{4}, \: \phi^\star(0) = \frac{\pi}{4}, \: Q= \mathbb{I}_2, \: R = 1$. During the traction phase, the line length and the kite velocity change as the line reels out. In our model, the parameter $\frac{v}{r}$ varies during operation and both the limit cycle and the dynamics around it would alter. The variations with respect to $\frac{v}{r}$ can be captured by modifying the periodic SE kernel for the multivariate case as described in Section \ref{sec:multivar}. The identification method with the extended kernel is applied on trajectory data from different operating conditions ($\frac{v}{r}\in\{0.3,0.2154,0.1625,0.1263,0.1\}$), where the training dataset consists of 16 trajectories with initial conditions randomly chosen from a uniform distribution with $||x_\perp(t_0)||_2 = 0.02$. Zero-mean Gaussian noise is added to the original state and state time-derivative measurements with an SNR of 60 dB. No disturbance is applied, i.e., $ d = 0$. The computation time in this example is around $1080\,$s.

Figure \ref{fig:kite_var} displays the identified models for two parameter values not used in training: Case 1: $\frac{v}{r}=0.11$ and Case 2: $\frac{v}{r} = 0.27$, with $A_{11}(\tau)$ and $g_2(\tau)$ as examples. The estimates $\hat{\Omega}(\tau)$ are very close to the analytical functions $\Omega(\tau)$. A trajectory is generated from Case 2 with an initial condition randomly chosen from a uniform distribution with $||x_\perp(t_0)||_2 = 0.1$. Figure \ref{fig:kite_var_traj_pred}(a) shows the predictions in the phase space of $\theta$ and $\phi$ using the identified model and a black-box kernel-SVM model trained with $(x,d,p)$ data as the input and $\dot{x}$ data as the output. The proposed method predicts the true nonlinear trajectory accurately, and performs significantly better than the black-box SVM method without the knowledge of the limit cycle. In Figure \ref{fig:kite_var_traj_pred}(b), the identified model is further compared with a model $\hat{\Omega}^{\text{med}}(\tau)$ identified only from the data at $\frac{v}{r}=0.1625$. The multivariate model clearly obtains better predictions than the model without $\frac{v}{r}$ parametrization.
\section{Conclusions}
A new methodology to identify the local limit cycle dynamics with an linear periodically parameter-varying model is presented. Decomposing the dynamics via transverse coordinates (done here by center surfaces) leads to linear periodic models that can accurately capture the local nonlinear dynamics around the limit cycle. The inherent periodicity is encoded in the identification in a non-parametric fashion by periodic kernels. This leverages the flexibility of kernel design by capturing, e.g., model variations due to changing operating conditions. Future research directions include: tailored kernel design and transversal surface selection approaches; applications to control design and large-scale problems, possibly in conjunction with model order reduction techniques. Other function learning algorithms such as neural network and online adaptation of the model are also interesting extensions to explore.

\bibliographystyle{IEEEtran}
\bibliography{IEEEabrv,bibliography}

\begin{thebibliography}{10}
\providecommand{\url}[1]{#1}
\csname url@rmstyle\endcsname
\providecommand{\newblock}{\relax}
\providecommand{\bibinfo}[2]{#2}
\providecommand\BIBentrySTDinterwordspacing{\spaceskip=0pt\relax}
\providecommand\BIBentryALTinterwordstretchfactor{4}
\providecommand\BIBentryALTinterwordspacing{\spaceskip=\fontdimen2\font plus
\BIBentryALTinterwordstretchfactor\fontdimen3\font minus
  \fontdimen4\font\relax}
\providecommand\BIBforeignlanguage[2]{{%
\expandafter\ifx\csname l@#1\endcsname\relax
\typeout{** WARNING: IEEEtran.bst: No hyphenation pattern has been}%
\typeout{** loaded for the language `#1'. Using the pattern for}%
\typeout{** the default language instead.}%
\else
\language=\csname l@#1\endcsname
\fi
#2}}

\bibitem{Strogatz1994}
S.~H. Strogatz, \emph{Nonlinear dynamics and chaos with applications to
  physics, biology, chemistry and engineering}.\hskip 1em plus 0.5em minus
  0.4em\relax Reading, Massachusetts: Addison-Wesley, 1994.

\bibitem{Schoukens2019}
J.~Schoukens and L.~Ljung, ``Nonlinear system identification: A user-oriented
  road map,'' \emph{IEEE Control Systems Magazine}, vol.~39, no.~6, pp. 28--99,
  2019.

\bibitem{toth2010modeling}
R.~T{\'o}th, \emph{Modeling and identification of linear parameter-varying
  systems}.\hskip 1em plus 0.5em minus 0.4em\relax Berlin, Heidelberg:
  Springer, 2010, vol. 403.

\bibitem{Allen2009}
M.~S. Allen and M.~W. Sracic, ``System identification of dynamic systems with
  cubic nonlinearities using linear time-periodic approximations,'' in
  \emph{7th International Conference on Multibody Systems, Nonlinear Dynamics,
  and Control}, vol.~4, 2009, pp. 731--741.

\bibitem{Manchester2011}
I.~R. Manchester, ``Transverse dynamics and regions of stability for nonlinear
  hybrid limit cycles,'' \emph{IFAC Proceedings Volumes}, vol.~44, no.~1, pp.
  6285--6290, 2011, 18th IFAC World Congress.

\bibitem{Bachnas2014}
A.~Bachnas, R.~T{\`o}th, J.~Ludlage, and A.~Mesbah, ``A review on data-driven
  linear parameter-varying modeling approaches: A high-purity distillation
  column case study,'' \emph{Journal of Process Control}, vol.~24, pp.
  272--285, 2014.

\bibitem{Laurain2012}
V.~Laurain, R.~Tóth, W.-X. Zheng, and M.~Gilson, ``Nonparametric
  identification of {LPV} models under general noise conditions: An {LS-SVM}
  based approach,'' \emph{IFAC Proceedings Volumes}, vol.~45, no.~16, pp.
  1761--1766, 2012, 16th IFAC Symposium on System Identification.

\bibitem{Rizvi2018}
S.~Z. Rizvi, J.~M. Velni, F.~Abbasi, R.~T{\`o}th, and N.~Meskin, ``State-space
  {LPV} model identification using kernelized machine learning,''
  \emph{Automatica}, vol.~88, pp. 38--47, 2018.

\bibitem{Diehl2013}
U.~Ahrens, M.~Diehl, and R.~Schmehl, \emph{Airborne Wind Energy}.\hskip 1em
  plus 0.5em minus 0.4em\relax Berlin, Heidelberg: Springer, 2013.

\bibitem{Hale1980}
J.~K. Hale, \emph{Ordinary Differential Equations}.\hskip 1em plus 0.5em minus
  0.4em\relax New York: R.E. Krierger Pub. Co., 1980.

\bibitem{Ahbe2022}
E.~Ahbe, A.~Iannelli, and R.~S. Smith, ``{A novel moving orthonormal
  coordinate-based approach for region of attraction analysis of limit
  cycles},'' \emph{Journal of Computational Dynamics}, 2022.

\bibitem{Toth2011}
R.~Tóth, V.~Laurain, W.~X. Zheng, and K.~Poolla, ``Model structure learning: A
  support vector machine approach for {LPV} linear-regression models,'' in
  \emph{50th IEEE Conference on Decision and Control and European Control
  Conference}, 2011, pp. 3192--3197.

\bibitem{SchoelkopfBook2002}
B.~Sch{\"o}lkopf, \emph{Learning with kernels: support vector machines,
  regularization, optimization, and beyond}, ser. Adaptive computation and
  machine learning.\hskip 1em plus 0.5em minus 0.4em\relax Cambridge,
  Massachusetts: MIT Press, 2001.

\bibitem{Alvarez2012}
M.~A. \'{A}lvarez, L.~Rosasco, and N.~D. Lawrence, ``Kernels for vector-valued
  functions: A review,'' \emph{Found. Trends Mach. Learn.}, vol.~4, no.~3, p.
  195–266, 2012.

\bibitem{Mackay1998}
D.~J. MacKay, ``Introduction to {Gaussian} processes,'' \emph{NATO ASI series
  F: computer and systems sciences}, vol. 168, pp. 133--166, 1998.

\bibitem{Rasmussen2006}
C.~E. Rasmussen and C.~K.~I. Williams, \emph{Gaussian processes for machine
  learning}, ser. Adaptive computation and machine learning.\hskip 1em plus
  0.5em minus 0.4em\relax Cambridge, Massachussets: MIT Press, 2006.

\bibitem{Wood2015}
T.~A. Wood, H.~Hesse, A.~U. Zgraggen, and R.~S. Smith, ``Model-based flight
  path planning and tracking for tethered wings,'' in \emph{54th IEEE
  Conference on Decision and Control (CDC)}, 2015, pp. 6712--6717.

\bibitem{Ahbe2018}
E.~Ahbe, T.~A. Wood, and R.~S. Smith, ``Stability verification for periodic
  trajectories of autonomous kite power systems,'' in \emph{European Control
  Conference (ECC)}, 2018, pp. 46--51.

\bibitem{Bittanti1991}
S.~Bittanti, P.~Colaneri, and G.~De~Nicolao, \emph{The Periodic Riccati
  Equation}.\hskip 1em plus 0.5em minus 0.4em\relax Berlin, Heidelberg:
  Springer, 1991, pp. 127--162.

\bibitem{Varga2007}
S.~Johansson, B.~K{\aa}gstr{\"o}m, A.~Shiriaev, and A.~Varga, ``Comparing
  one-shot and multi-shot methods for solving periodic {Riccati} differential
  equations,'' in \emph{IFAC Proceedings Volumes}, vol.~3, no.~1, 2007, pp.
  163--168.

\end{thebibliography}

\end{document}